\documentstyle[pra,twocolumn,psfig,aps]{revtex} 

\begin{document}
\draft
\def\overlay#1#2{\setbox0=\hbox{#1}\setbox1=\hbox to \wd0{\hss #2\hss}#1
-2\wd0\copy1}

\twocolumn[\hsize\textwidth\columnwidth\hsize\csname@twocolumnfalse\endcsname

\title{Using Conditional Measurements to Combat Decoherence}
\author{M.~Fortunato$^{*(a)}$ \and G.~Harel$^{(b)}$ and G.~Kurizki$^{(b)}$}
\address{(a) Dipartimento di Matematica e Fisica, Universit\`a di Camerino
         I--62032 Camerino
          (MC) Italy \\ and Istituto Nazionale per la Fisica della 
          Materia, Unit\`a di Camerino, Italy}
\address{ (b) Dept. of Chemical Physics, The Weizmann Institute of Science,
 Rehovot 76100, Israel}
\date{\today}
\maketitle
\begin{abstract}
With the help of some remarkable examples, it is shown that conditional
measurements performed on two-level atoms just after they have interacted
with a resonant cavity field mode are able to recover the coherence of
number-state superpositions, which is lost due to dissipation. 
\end{abstract}
\vskip3pc]
\narrowtext

\section{Introduction: Decoherence of Non-Classical States}
\label{intro}

The phenomenon of decoherence~\cite{kn:zurek} is a fundamental aspect of the dynamics of
open quantum systems, since it rapidly destroys the characteristic feature of non-classical
states of a quantum ``object'', the so-called quantum coherence between the components of a
superposition state, leading to the corresponding classical mixture.
Recently, decoherence has also acquired a great applied importance, because it determines the
feasibility of quantum information storage, encoding (encrypting) and computing~\cite{kn:qc}.
Many proposals have been made in the last years to combat the irreversibility of decoherence 
processes in quantum computing.
One consists of the filtering out of the part of the ensemble which has not
decohered~\cite{kn:yama}. The other proposal is encoding the state (qubit) by means of several
ancillas,  decoding the result after a certain time, checking the ancillas for  error syndromes
and correcting them~\cite{kn:others}. With the latter approach, only extremely small error
probabilities can be addressed~\cite{shordoria}.
 
The essence of quantum computation demands that decoherence be countered without 
the knowledge of which state is in error during the computation. However, there is
a simpler but still important problem: how to protect from decoherence
the {\it input} states, before the beginning of the actual computation.
We propose here an approach which is not unitary, but
which can be safely used to store quantum field states in dissipative cavities,
in order to successively use them as inputs in information 
processing or in signal transmission.
The basic idea consists of {\it restoring} the decohered field state 
by letting it interact with a two-level atom, and then projecting the
resulting entangled state onto a superposition of atomic eigenstates,
whose phase and moduli are determined as a function of the field state
one desires to recover.
Such projection is accomplished by a conditional measurement (CM)~\cite{kn:cm,kn:ens} of the
appropriate atomic state~\cite{kn:gil}, in the context of the Jaynes-Cummings (JC) model.
In the present problem we set the initial (unspoilt) superposition of 
zero-photon up to $N$-photon states as our target state,
and work in Liouville space instead of Hilbert space,
so as to account for the state decoherence.
The results demonstrate that a few {\it highly-probable} CMs, 
in this simple model, can drastically reduce even a large error.
One of our objectives is to find the optimal compromise between the CM success
probability and the error size, which grows in the course of dissipation. 

The ability to {\it approximately restore any mixture to any pure state} 
is the advantage of our post-selection CM approach,
compared to the non-selective measurement (tracing) approach.

\section{Recovering Coherence by Optimal Conditional Measurements}
\label{dmbcm}

Let us consider a single-mode cavity in which the quantized electromagnetic field 
is initially prepared in a finite superposition of Fock states, 
\begin{equation}
|\phi(0)\rangle = \sum_{n=0}^N c_n|n\rangle \;.
\label{eq:infieldstate}
\end{equation}
We assume that the cavity field is in interaction with a (zero-temperature, for the sake of
simplicity) heat bath, to take into account the effect of dissipation. The resulting master
equation describing such coupling, in the interaction picture, is
\begin{equation}
\dot{w}_{_F}  =  \gamma(2\hat{a}w_{_F}\hat{a}^\dagger-\hat{a}^\dagger
\hat{a}w_{_F}-w_{_F}\hat{a}^\dagger\hat{a})\;,
\label{eq:meq}
\end{equation}
where $w_{_F}(t)$ is the density operator of the field mode,
$\hat{a}$ and $\hat{a}^\dagger$ are the annihilation and creation operators 
of the field, and $\gamma$ is the damping constant of the cavity. 

The solution of Eq.~(\ref{eq:meq}) after dissipation over 
time $\bar{t}>0$ \cite{kn:arnoldus} is given by
\begin{eqnarray}
w_{n,m}(\bar{t}) & = & \sum_{k=0}^{\infty} 
 w_{n+k,m+k}(0)
 \sqrt{{\scriptsize\pmatrix{n+k\cr n} }
 (e^{-2\gamma\bar t})^n 
 (1-e^{-2\gamma\bar t})^k }
 \nonumber \\ & & \times 
 \sqrt{{\scriptsize\pmatrix{m+k\cr m} }
 (e^{-2\gamma\bar t})^m 
 (1-e^{-2\gamma\bar t})^k }
 \;,
\label{eq:solro}
\end{eqnarray}
written here in Fock basis,
$w_{n,m}(t)=\langle n|w_{_F}(t)|m\rangle$. 

In order to recover the original state of the field
we propose to apply 
an optimized CM (or a sequence thereof) to the cavity: 
Using a classical field we prepare a two-level atom in a chosen 
superposition~\cite{kn:gil,kn:meschede}
\begin{equation}
|\psi^{(i)}\rangle = \alpha^{(i)} |a\rangle + \beta^{(i)}|b\rangle
\label{eq:inatstate}
\end{equation}
of its ground $|b\rangle$ and excited $|a\rangle$ states, 
and let it interact with the field for a time $\tau$ by sending it through the 
cavity with controlled speed.
The field-atom interaction is adequately described by the resonant 
Jaynes-Cummings (JC) model~\cite{kn:jc}.
We assume the field-atom interaction time $\tau$ to be much shorter than the
cavity lifetime, $\gamma\tau \ll 1$,
so that we are allowed to neglect dissipation during each CM.
Upon exiting the cavity the atom is {\it conditionally measured},
using a second classical field, to be in a state 
\begin{equation}
|\psi^{(f)}\rangle = \alpha^{(f)} |a\rangle +
                            \beta^{(f)}|b\rangle\;,
\label{eq:finatstate}
\end{equation}
which is in general different from the initial atomic state $|\psi^{(i)}\rangle$. 
This means that we post-select, using the same setup as in Ref.~\cite{kn:gil},
the atomic superposition state~(\ref{eq:finatstate}) which is {\it correlated}
to a cavity field state that is as close as possible to the original 
state~(\ref{eq:infieldstate}).

The effect of the applied CM on the cavity field is then calculated as follows:
Initially, at the time the atom enters the cavity, 
the density matrix of the field-atom system is 
\begin{equation}
w_{_{FA}}(\bar{t})=
 w_{_F}(\bar{t}) \otimes |\psi^{(i)}\rangle\langle\psi^{(i)}|\;.
\label{eq:intotdens}
\end{equation}
It then evolves unitarily by the JC interaction of duration $\tau$ into 
\begin{equation}
w_{_{FA}}(\bar{t}+\tau)=\hat{U}(\tau)
w_{_{FA}}(\bar{t})\hat{U}^\dagger(\tau)\;,
\label{eq:evolvrho}
\end{equation}
where $\hat{U}(\tau)$ is the (interaction picture) time evolution operator
\begin{eqnarray}
\hat{U}(\tau) |n\rangle|a\rangle & = & 
 C_n |n\rangle|a\rangle -iS_n |n+1\rangle|b\rangle 
\label{eq:uneva} \nonumber\\
\hat{U}(\tau) |n\rangle|b\rangle & = & 
 C_{n-1} |n\rangle|b\rangle -iS_{n-1}|n-1\rangle|a\rangle\; ,
\label{eq:unevb}
\end{eqnarray}
where $C_n=\cos\left(g\tau\sqrt{n+1}\right)$ and 
$S_n=\sin\left(g\tau\sqrt{n+1}\right)$,
and $g$ equals the field-atom coupling strength, {\it i.e.} the vacuum Rabi 
frequency.
Finally, the conditional measurement of the atom in the state 
$|\psi^{(f)}\rangle$ results in a density matrix of the {\it field} given by 
\begin{equation}
w_{_F}(\bar{t}+\tau)={\rm Tr}_{_A} \left[w_{_{FA}}(\bar{t}+\tau)
|\psi^{(f)}\rangle\langle\psi^{(f)}|\right]/P\;,
\label{eq:traceformula}
\end{equation}
where
\begin{equation}
P={{\rm Tr}_{_{F}}{\rm Tr}_{_A}\left[w_{_{FA}}(\bar{t}+\tau)
|\psi^{(f)}\rangle\langle\psi^{(f)}|\right]}
\label{eq:probability}
\end{equation}
is the success probability of the CM. 

In order to approximately recover the initial state of the field, we use the dependence of 
$w_{_{F}}(\bar{t}+\tau)$ on the initial and final atomic states and the
field-atom interaction time,
choosing {\it optimal} parameters $\alpha^{(i)}$, $\beta^{(i)}$, 
$\alpha^{(f)}$, $\beta^{(f)}$ and $\tau$ such that the relation
\begin{equation}
w_{_{F}}(\bar{t}+\tau) \approx w_{_{F}}(0) 
\label{rc}
\end{equation}
is satisfied.
These optimal CM parameters are found by minimizing the cost 
function~\cite{kn:gil} 
\begin{equation}
G=\frac{d(w_{_{F}}(\bar{t}+\tau),w_{_{F}}(0))}{P^r}\;, 
\label{eq:cf} \end{equation}
where $d$ is a distance function between two density operators, defined as
\begin{equation} d(w_{_F}^{(1)},w_{_F}^{(2)})=
 \sqrt{\sum_{nm}(w^{(1)}_{nm}-w^{(2)}_{nm})^2}\;,
\label{eq:df}\end{equation}
$P$ is the CM success probability (\ref{eq:probability}),
and the tunable exponent $r>0$ determines the relative importance of the two 
factors in $G$.
If this CM does not bring us as close to the original state as our experimental
accuracy permits, we can repeat the process over and over again, as long
as the distance to the original state keeps decreasing, while the CM success
probability remains high.  The specific form of the atomic states (\ref{eq:inatstate}) and 
(\ref{eq:finatstate}) is chosen by {\it minimizing} Eq.~(\ref{eq:cf}) at each step.
We emphasize that the application of each CM may introduce
widening of the photon-number distribution by one photon,
and yet the {\it optimized} CMs are able of preventing this widening and,
moreover, of restoring the field to its initial pure state.
This ability amounts to an effective control of a large Fock-state 
subspace.

\section{Example}
\label{example}

In this section we illustrate our proposal with the help of an example,
making use of the Husimi $Q$-function defined as
$ Q_{w_{_F}}(\beta,\beta^*)= \langle\beta|w_{_F}|\beta\rangle$,
where $|\beta\rangle$ represents a coherent state of complex amplitude $\beta$, 
to display the error-correction process.

Let us take as the original field state a symmetric
superposition of the vacuum and one-photon state,
\begin{equation}
|\phi(0)\rangle=\frac{1}{\protect\sqrt{2}}(|0\rangle+e^{i\pi/3}|1\rangle)\;,
\end{equation}
whose $Q$-function is shown in Fig.~1(a).
Dissipation by $\gamma \bar{t}=0.3$ makes the {\it error} matrix
$ w_{_F}(\bar t)-w_{_F}(0) $ of considerable magnitude,
as seen in Fig.~1(b).
After the application of one CM
($|\psi^{(i)}\rangle =\cos(3\pi/8)|a\rangle + 
\sin(3\pi/8)e^{i5\pi/4}|b\rangle$,
 $g\tau=37.95$,
 $|\psi^{(f)}\rangle =\cos(3\pi/8)|a\rangle + \sin(3\pi/8)e^{i\pi/4}|b\rangle$),
 optimized to yield high success probability
($r=2$), the resulting error matrix 
$ w_{_F}(\bar t+\tau)-w_{_F}(0) $ is reduced by a factor of about 3, as
it is visible in Fig.~1(c). 
The success probability of the CM is markedly high (74\%).
Subsequent CMs can further reduce the distance to $1/6$ (one sixth) its original
magnitude, with 62\% success probability for the full CM sequence.
Stronger error reduction is obtainable at the expense of success probability: 
the application of 4 CMs optimized for $r=1$ (respectively $r=0$) yields an 
error reduction factor of 11 (respectively 28) with sequence probability of 
33\% (respectively 16\%).

We emphasize that the application of our approach is not limited to 
equal-amplitude superposition states: indeed the error correction is
even better achieved for strongly unequal superpositions.

A full analysis of the distance $d_K=d(w^K,w_{_F}(0))$ [Eq.~(\ref{eq:df})]
between the recovered state and the original state
and of the CM sequence probability $P_{seq,K}=\prod_{l=1}^K P_l$,
with $P_l$ given by~(\ref{eq:probability}), as a function
of the number of CMs performed shows that the first CMs achieve
a strong reduction of such a distance, whereas after a few successive CMs
saturation sets on, in terms of both distance and success probability.

It is very interesting to note that the success probability in our 
approach is often larger (and sometimes even much larger) than the
{\it theoretical} probability to find the original state
in the dissipation-spoilt state, namely,
${\rm Tr}_{_F}[w_{_F}(0)w_{_F}(\bar t)]$,
which we call the {\it filtering probability}.

\section{Conclusions}
\label{conclu}

We have  shown the ability of simple JC-dynamics CMs as an effective means
of reversing the unwanted effect of dissipation 
on coherent superpositions of Fock-states of a cavity field: 
the successive application of a small number of optimized CMs 
recovers the original (pure) state of the field with high success probability,
which {\it is comparable or even surpasses} the filtering probability. 
The simplest tactics may employ a {\it single highly-probable trial} 
to achieve nearly-complete error correction.
Surprisingly, even though we have only five control parameters at our
disposal for each CM, our optimization procedure is able to effectively 
control the amplitudes in a large Fock-state subspace~\cite{kn:mau}.

Among the practical difficulties an experimenter might encounter in the
application of any CM approach~\cite{kn:gil,kn:kozh,kn:gil2},
realistic atomic velocity fluctuations (of 1\%) and cavity-temperature
effects (below $1^\circ$K) are relatively unimportant,
and especially so in the present scheme which makes use of a single or few CMs
so that the effect of experimental imperfections is linear in the input errors.
Only atomic detection efficiency is an experimental challenge~\cite{kn:gil2}.
Although the detection efficiency is currently low, it is expected to rise
considerably in the coming future.

In conclusion, we would like to stress that extensions of this approach to
more complex field-atom interaction Hamiltonians can make this correction
procedure effective even with fewer CMs and for highly complicated states, 
encoding many qubits of information.
However, even in its current simple form the suggested approach has undoubted
merits:
(i) it can yield higher success probabilities than the filtering approach;
(ii) it is not limited to small errors as ``high level'' unitary-transformation
approaches are;
(iii) it corrects errors after their occurrence,
with no reliance on ideal continuous monitoring of the dissipation
channel and on instantaneous feedback; and
(iv) it is realistic in that it can counter combined phase-amplitude errors
which arise in cavity dissipation, and is of general applicability, that is,
it is not restricted to specific models of dissipation.

\acknowledgments
\label{acknow}

We acknowledge the support of the German-Israeli Foundation (GIF).
One of us (MF) thanks the European Union (TMR programme) and INFM
(through the 1997 Advanced Research Project ``CAT'') for support.

\begin{figure}
{\centerline{\vbox{
\psfig{file=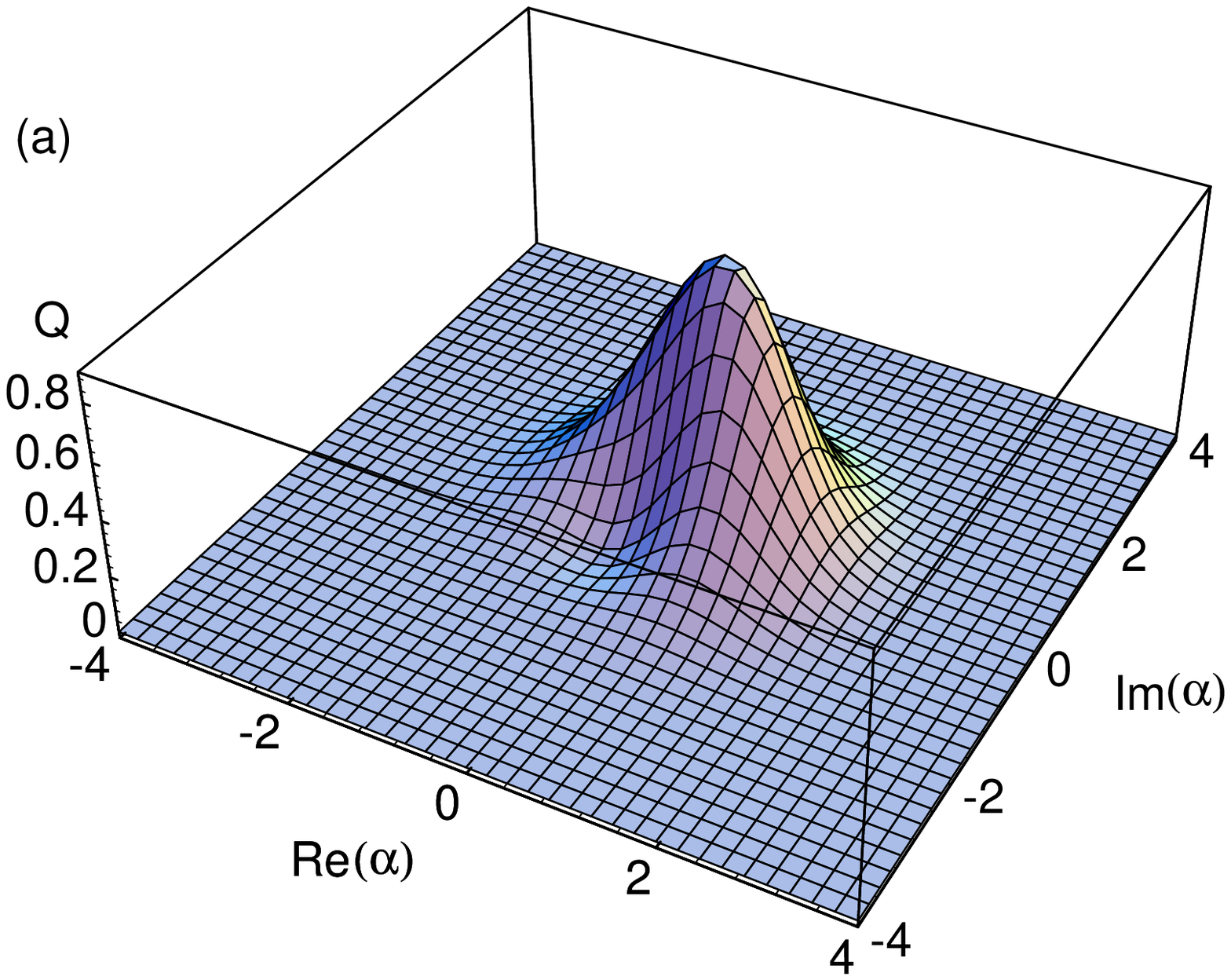,width=7.5cm,angle=00}
\psfig{file=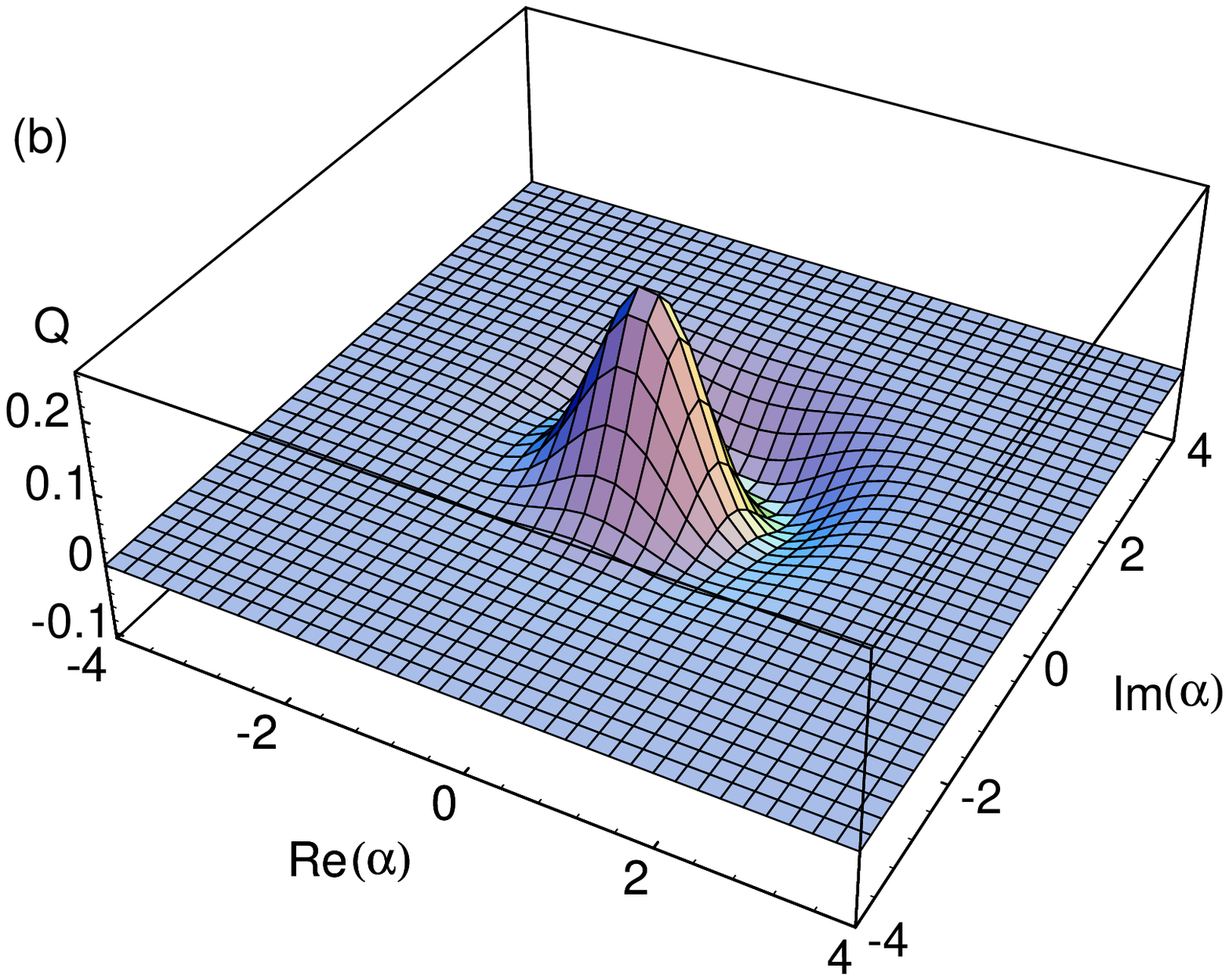,width=7.5cm,angle=00}
\psfig{file=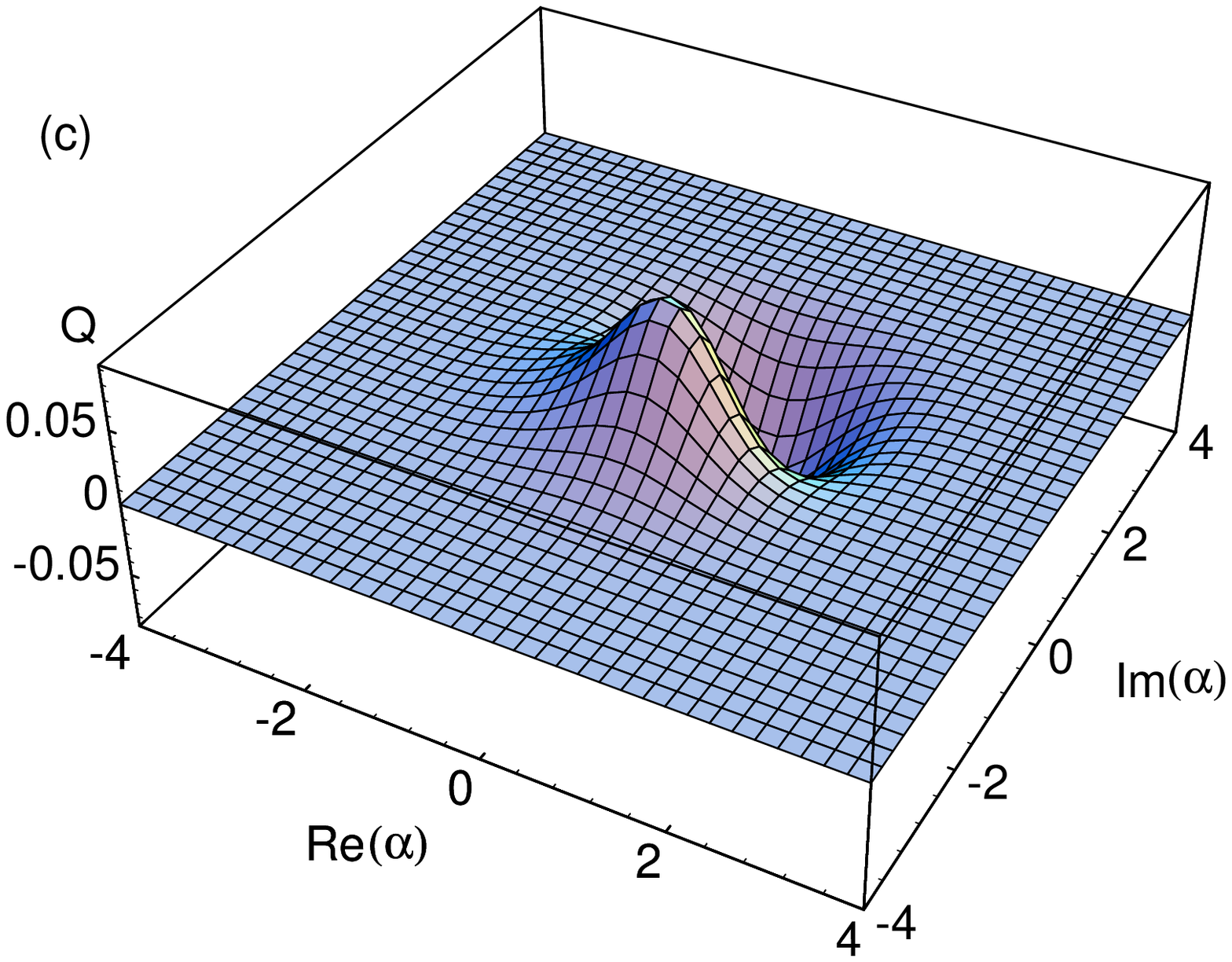,width=7.5cm,angle=00}
}}}
\vskip .3 cm
\caption{
$Q$-function description of
(a) original field, with $w_{_F}(0)=|\phi(0)\rangle\langle\phi(0)|$,
$|\phi(0)\rangle=(|0\rangle+e^{i\pi/3}|1\rangle)/\protect\sqrt2$;
(b) error after dissipation, $w_{_F}(\bar t)-w_{_F}(0)$;
(c) reduced error after 1 optimized CM [minimizing Eq.~(\protect\ref{eq:cf})],
 $w_{_F}(\bar t+\tau)-w_{_F}(0)$.
}
\end{figure}

\end{document}